\documentclass[aps,pre,twocolumn,groupedaddress,showpacs,amsmath,amssymb]{revtex4}
\usepackage{graphicx}

\begin{document}
\title{Ehrenfest time dependence of quantum transport corrections and spectral statistics}
\author{Daniel Waltner and Jack Kuipers}
\affiliation{Institut f\"ur Theoretische Physik, Universit\"at Regensburg, D-93040 Regensburg, Germany}
\begin{abstract}
The Ehrenfest time scale in quantum transport separates essentially classical propagation from wave interference and here we consider its effect on the transmission and reflection through quantum dots.  In particular we calculate the Ehrenfest time dependence of the next to leading order quantum corrections to the transmission and reflection for dc- and ac-transport and check that our results are consistent with current conservation relations. Looking as well at spectral statistics in closed systems, we finally demonstrate how the contributions analyzed here imply changes in the calculation given in [P.\ W.\ Brouwer, S.\ Rahav and C.\ Tian, Phys.\ Rev.\ E {\bf 74}, 066208 (2006)] of the next to leading order of the spectral form factor.  Our semiclassical result coincides with the result obtained in [C.\ Tian and A.\ I.\ Larkin, Phys. Rev. B {\bf 70}, 035305 (2004)] by field-theoretical methods.
\end{abstract}
\maketitle

\section{Introduction}

Chaotic quantum systems are expected \cite{Boh} to show universal behavior that can be described by Random Matrix Theory (RMT) \cite{Met}. After this was conjectured \cite{Boh} the challenge was to justify and dynamically understand the relation between chaotic systems and RMT. Here semiclassical methods have proved to be very successful \cite{Ber1,Han,Sie,Mul1,Heu}. These are based on asymptotic expansions of the quantum propagator, the Green function and its trace, which consist of sums over classical trajectories \cite{Gut}.  Each sum contains, along with prefactors determined by the classical dynamics, phases determined by the classical actions of the trajectories which allow for interference effects.

On one front, these semiclassical methods were applied to study spectral properties of {\it closed} systems. Here one considers for example the spectral autocorrelation function
\begin{equation}
\label{in1}
{\hat K}(\omega) =2\pi\hbar\left\langle\rho_{\rm osc}\left(E+\hbar\omega/2 \right) \rho_{\rm osc}\left(E-\hbar\omega/2 \right)\right\rangle,
\end{equation}
defined as the energy averaged correlation function of the oscillating parts of two densities of states $\rho_{\rm osc}(E)$ with an energy difference $\hbar\omega$. Here and in the following $\left\langle\ldots\right\rangle$ denotes an average over a classically small but quantum mechanically
large energy window $\Delta E$. Semiclassics now enters by replacing the spectral densities by their semiclassical expression in terms of a sum over periodic orbits given by the Gutzwiller trace formula \cite{Gut}
\begin{equation}
\label{in2}
\rho_{\rm osc}\left(E \right)\sim\Re\sum_\gamma A_\gamma{\rm e}^{(i/\hbar) S_\gamma(E)},
\end{equation}
for $\hbar\rightarrow 0$ with $A_\gamma$ the stability amplitudes (for their exact form see \cite{Gut} for example) and $S_\gamma(E)$ the classical actions of the periodic orbits and $\Re$ denoting the real part.  Linearizing the actions around the energy $E$ finally yields when defining
\begin{eqnarray}
\label{in3}
K(\omega)&\sim&2\pi\hbar\left\langle\sum_{\gamma\gamma'} A_\gamma A_{\gamma'}^*{\rm e}^{(i/\hbar) \left( S_\gamma(E)-S_\gamma'(E)\right)}\right.\nonumber\\ &&\left.\times{\rm e}^{(i\omega/2)\left(T_\gamma(E)+T_{\gamma'}(E)\right)}\right\rangle
\end{eqnarray}
for
\begin{equation}
{\hat K}(\omega)\sim2\Re K(\omega)
\end{equation}
with $T_\gamma$ the period of the orbit $\gamma$.
Often the spectral form factor $K(\tau)$, which is Fourier transform of $K(\omega)$, is considered
\begin{equation}
\label{in4}
K(t)=\frac{1}{2\pi}\int d\omega{\rm e}^{-i\omega t}K(\omega).
\end{equation}

\begin{figure}
\begin{center}
\includegraphics[width=8cm]{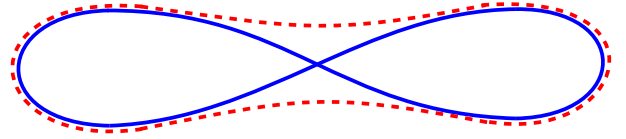}
\caption{\label{fig0} Correlated orbits analyzed by Sieber and Richter \cite{Sie} which differ in an encounter and lead to the first off-diagonal correction to the spectral form factor.}
\end{center}
\end{figure}

Importantly, the expressions for $K(\omega)$ and $K(\tau)$ oscillate rapidly depending on $E$; dominant contributions will thus result from trajectories with very similar actions. For example, the diagonal contribution, i.e.\ $\gamma=\gamma'$ with equal action, was first studied in 1985 in \cite{Ber1}, using the sum rule of \cite{Han}. It yielded the leading order RMT-prediction in $1/(i\omega)$ for the spectral autocorrelation function and thus the leading order in $\tau$ for the spectral form factor. Off-diagonal contributions were first taken into account by Sieber and Richter in 2001 \cite{Sie} by considering two orbits essentially differing from each other in a encounter region where the two orbits are differently connected, see Figure~\ref{fig0}.
This work could later be extended and formalized yielding the RMT results to arbitrary high order in powers of $\tau$ \cite{Mul1,Heu} and also extended to other quantities characterizing the spectral properties of a system \cite{Wal}.

On another front, for {\it open} systems the conductance was analyzed semiclassically within the Landauer-B\"uttiker approach \cite{Lan} to transport.  In particular we imagine a system connected to two leads carrying $N_1$ and $N_2$ channels respectively (with a total $N=N_1+N_2$).  The conductance is related to the transmission matrix elements $t_{\alpha,\beta}$ via the Landauer-B\"uttiker approach and these matrix elements can be approximated semiclassically (for an overview see \cite{Wal1}) by 
\begin{equation}
\label{in5}
t_{\alpha,\beta}\sim\frac{1}{\sqrt{T_H}}\sum_{\gamma(\beta\to\alpha)}B_\gamma{\rm e}^{\left(i/\hbar \right) \left( S_\gamma(E)\right)}
\end{equation}
with $t_{\alpha,\beta}$ characterizing the transitions between the modes $\beta$ and $\alpha$, and $\gamma$ the classical scattering trajectories that connect those modes in the two different leads. The Heisenberg time is defined as $T_H=2\pi\hbar\overline{d}(E)$ with $\overline{d}(E)$ the mean spectral density of the considered closed system. For the exact form of the stability amplitudes $B_\gamma$ and classical actions $S_\gamma(E)$ see e.g.\ \cite{Wal1}. A corresponding expression in terms of trajectories connecting one lead to itself holds for reflection amplitudes $r_{\alpha,\beta}$. Using eq.\ (\ref{in5}) we obtain for the transmission $T$ characterizing the conductance
\begin{equation}
\label{in6}
T\equiv \left\langle {\rm Tr}\left(tt^\dagger\right)\right\rangle\sim \frac{1}{T_H}\left\langle \sum_{\gamma\gamma'}B_\gamma B_{\gamma'}^*{\rm e}^{\left(i/\hbar \right) \left( S_\gamma(E)-S_{\gamma'}(E)\right)}\right\rangle. 
\end{equation}
The sum runs over all paths $\gamma$ and $\gamma'$ which connect the two leads.  A similar expression also holds for the reflection $R\equiv \left\langle {\rm Tr}\left(rr^\dagger\right)\right\rangle$.   To be precise we will use $R$ to denote the reflection into lead 1 in the following, but the expression for the reflection into lead 2 just follows by swapping $N_1$ and $N_2$. The first nondiagonal (i.e.\ next order in inverse channel number $1/N$) contribution for the conductance was analyzed in \cite{Ric} yielding again a result consistent with RMT. The extension of the conductance to arbitrary high order was performed in \cite{Heu1} and the authors later treated the shot noise \cite{Bra} and other correlation functions like the conductance variance in \cite{Mul2}.  For this the authors built on their work on closed systems \cite{Mul1} and noticed in particular that the diagrams of correlated pairs of scattering trajectories that appear for the conductance can be created by cutting the pairs of periodic orbits that contribute to the spectral form factor once and moving the cut ends to the leads.  Likewise, the diagrams of trajectory quadruplets that appear for the conductance variance can be obtained by cutting the periodic orbit pairs exactly twice.

Up to now we only discussed one application of these semiclassical techniques, the confirmation of RMT-results. However it is also possible to predict effects away from this arena, i.e.\ the behavior of chaotic systems for a finite Ehrenfest time. The Ehrenfest time $\tau_E=(1/\lambda) \ln (E/(\lambda\hbar))$; more generally defined as a time proportional to $\ln\hbar$ \cite{Chi}; is the time needed for a wave-packet to reach a size such that it can no longer be described as a single classical particle. The Ehrenfest time thus separates the free evolution of wave packets that follow essentially the classical dynamics from the evolution on larger time scales where wave interference becomes dominant.
Including the Ehrenfest time into the calculation of the quantities presented above started with the pioneering work \cite{Ada} that calculated the first quantum correction to the energy-averaged transmission. This analysis was extended to reflection \cite{Jac,Bro} including also a distinction between different Ehrenfest times \cite{Jac}.  An exponential suppression of this quantum correction proportional to ${\rm e}^{-\tau_E/\tau_D}$ was observed involving the dwell time $\tau_D$, the average time the particle stays inside an open billiard. Furthermore the Ehrenfest time dependent behavior of other transport quantities soon followed: the independence of (the leading order of) the universal conductance fluctuations was obtained in \cite{Bro}, the shot noise and Fano factor were found to be exponentially suppressed like the averaged transmission in \cite{Whi}, and the behavior of a third order correlation function was derived in \cite{Bro1}.

\begin{figure}
\begin{center}
\includegraphics[width=8cm]{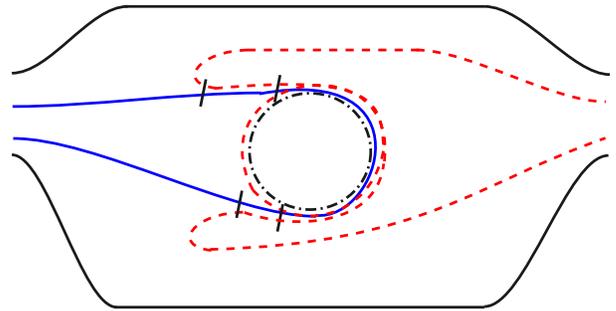}
\caption{\label{fig1} Diagram occurring in the calculation of the reflection covariance (or the conductance variance) containing two orbits surrounding a central periodic orbit. The fringes are marked by (black) vertical lines perpendicular to the trajectories. Partner orbits are not shown.}
\end{center}
\end{figure}

Of these it is the treatment of the conductance variance \cite{Bro} we are particularly interested in here.  Because of the unitarity of the scattering matrix this is equal to the reflection covariance, which turns out to be slightly simpler to treat semiclassically, and the authors found one important contribution was given by a diagram like in Figure~\ref{fig1}.  There two trajectories (one from either lead) approach a trapped periodic orbit with one winding around it an extra time.  Partner trajectories (not shown) can be found which follow those trajectories almost exactly but where one winding is exchanged between the two trajectories 
leading to a quadruplet of trajectories with a small action difference and a contribution in the semiclassical limit.  Such a contribution vanishes when the Ehrenfest time goes to 0 and can be seen to contain the discrete diagram types considered previously without Ehrenfest time \cite{Mul2} (which then naturally sum to 0).  Although this contribution vanishes, similar periodic orbit encounters can contribute in other situations \cite{Kui,Wal} when the Ehrenfest time is 0.

Combined with another diagram which does not involve a periodic orbit encounter, \cite{Bro} showed the independence of the conductance variance of the Ehrenfest time.  Later these techniques were applied to the spectral autocorrelation function and the spectral form factor for closed systems in \cite{Rah}.  This allowed the authors of \cite{Rah} to obtain the first quantum correction in the case with and without time reversal symmetry, but they found a discrepancy with the field-theoretical result \cite{Tia} obtained using effective RMT, a phenomenological approach to mimic the Ehrenfest time behavior in the RMT framework. We investigate this discrepancy here in this article and show how a hierarchy of diagram possibilities (e.g.\ reversing the cutting of periodic orbits to create diagrams) restores the consistency in the semiclassical treatment.

Currently all the Ehrenfest time approaches (described above) are restricted to very low order in the inverse channel number for the transmission and in $1/\omega$ or $\tau$ for the spectral autocorrelation function or form factor. A calculation of the corrections to infinite order, as has been performed in the case of vanishing Ehrenfest time $\tau_E/\tau_D\rightarrow 0$, is still lacking.  We want in this paper to make a step towards filling this gap. More precisely we consider in the section II the next-to-leading order quantum correction to the transmission and reflection in the case of the dc-transport with and without time reversal symmetry. We then check the unitarity of our result, i.e.\ that $T$ and $R$ add up to a constant ($N_1$) at the considered order. In section III we extend the results of section II to ac-transport and then check that corrections to the closely related Wigner time delay are indeed zero at the order considered. In section IV we apply our previous results to closed systems to obtain for the spectral form factor with Ehrenfest time a result consistent with the field-theoretical prediction \cite{Tia} and finally conclude.

\section{Transmission and reflection}

Before we turn to the form factor later we remain with quantum transport and consider the transmission and reflection. The leading order contribution (in inverse channel number $1/N$) to the transmission and reflection results from the diagonal approximation (pairing $\gamma=\gamma'$ in (\ref{in6})).  The calculation of this contribution can be found for example in \cite{Bar} and the result is independent of the Ehrenfest time and is of order $N$. The next order in inverse channel number (i.e.\ of order 1) results from the periodic orbit pairs shown in Figure~\ref{fig0}, where one of the two loops (that which is traversed in the same direction by both orbits) is cut open and the two ends are brought to the two openings \cite{Ric,Heu1} and its contribution is damped exponentially with the Ehrenfest time \cite{Ada}. For the reflection an additional possibility arises, called coherent backscattering, and which can be created by cutting the orbits in Figure~\ref{fig0} in half (keeping the half traversed in different directions by the orbits) and moving what is left of the encounter to the lead.  As there is still the remnant of the encounter, this case is also suppressed exponentially with the Ehrenfest time \cite{Jac,Bro}.  This dependence is essential for the unitarity of the scattering so that if we sum the transmission and the reflection these off-diagonal corrections cancel.  Of course as both involve a closed loop which is traversed in two different directions by the trajectory and its partner, they do not exist and can yield no contribution when time reversal symmetry is absent.

But it is the next order contributions we are particularly interested in, and we start with the simpler case where the scattering system does not have time reversal symmetry.

\subsection{No time reversal symmetry}
\label{unitrans}

\begin{figure}
\begin{center}
\includegraphics[width=8.6cm]{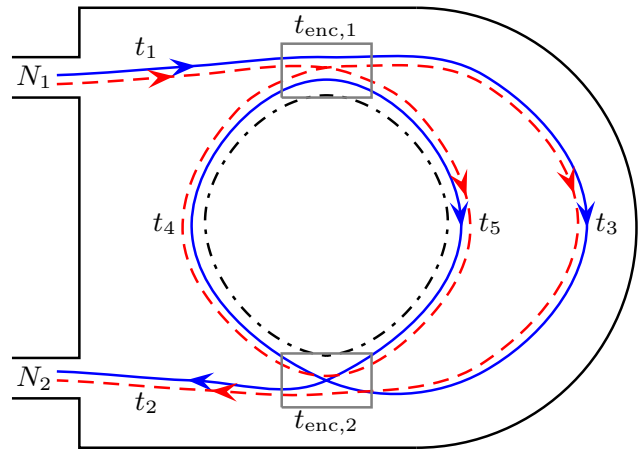}
\caption{Example of an orbit (and its partner shown dashed), considered in \cite{Heu1}, that contributes to the transmission for systems without time reversal symmetry.  A central periodic orbit (dashed-dotted) can be identified.}
\label{fig2,5}
\end{center}
\end{figure}
\begin{figure}
\begin{center}
\includegraphics[width=8cm]{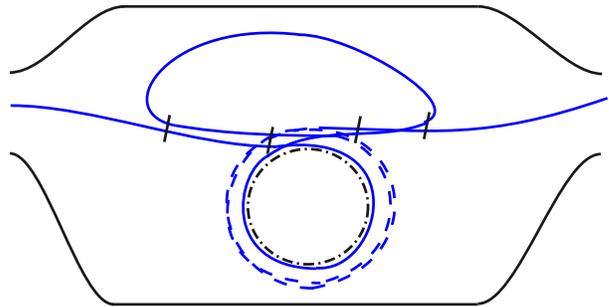}
\caption{Diagram studied in the calculation of the first quantum correction to the transmission and reflection in the absence of time reversal symmetry. In this example, the orbit traverses the central periodic orbit (dashed-dotted line) once during its first encounter and twice during its second encounter.  We draw the parts of the orbit during the second encounter dashed to distinguish them from the first.  The partner orbit (not shown) has one traversal of the central periodic orbit exchanged between its first and the second encounter with the periodic orbit (i.e.\ it goes around twice then once). The fringes are marked by black vertical lines perpendicular to the trajectories.}
\label{fig2}
\end{center}
\end{figure}
\begin{figure}
\begin{center}
\includegraphics[width=8.6cm]{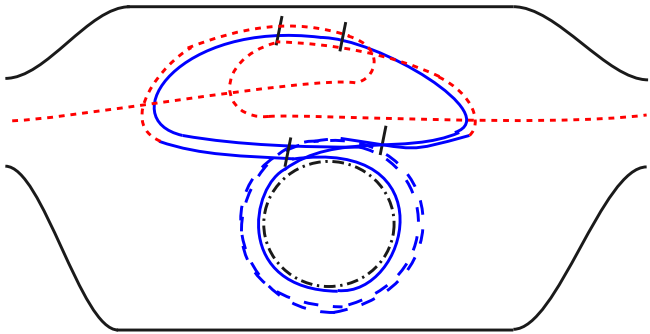}
\caption{Example of an orbit, that has to be considered in the calculation of the transmission, however not in the case of the calculation of the conductance fluctuations. The parts of the orbit that were changed in comparison to Fig.\ \ref{fig2} are shown dotted (red).}
\label{fig3}
\end{center}
\end{figure}

The $1/N$ order contribution results from orbits with two encounters with itself \cite{Heu1}, see Fig.~\ref{fig2,5}. We can see that there is a central periodic orbit through the two encounters.  This fact is essential for the Ehrenfest time dependence and simplifies treating the different cases.  Depending on how much these encounters overlap (i.e.\ depending on the lengths of the links $t_4$ and $t_5$ in Fig.~\ref{fig2,5}), one distinguishes in the case of no overlap two independent 2-encounters (i.e.\ encounters involving 2 orbit stretches), in the case the two 2-encounters overlap at one of their ends (shrinking $t_4$ or $t_5$ say) a 3-encounter and in the case the two 2-encounters overlap at both ends (shrinking $t_4$ and $t_5$) an encounter fully surrounding the periodic orbit. Although we mentioned up to now only one contained periodic orbit shown dashed-dotted in Fig.~\ref{fig2,5}, there are two in total: one built up by $t_4$ and $t_5$, the other by $t_3$ and $t_4$. In the following calculations we choose either as we actually treat this configuration as an orbit meeting a central periodic orbit twice, see Fig.~\ref{fig2}. This procedure counts every configuration twice; this overcounting factor accounts for the fact that for fixed orbit parts, i.e.\ for fixed dashed-dotted periodic orbit and fixed orbit encountering it in Fig.~\ref{fig2}, we have two possibilities to construct an orbit pair: the original orbit can surround the dashed-dotted orbit once more than its partner during either the first or the second encounter. The two possibilities correspond to swapping the original orbit and its partner, 
and both terms are included in the sum over orbits in (\ref{in6}).  The equivalence between 
chosing a different central periodic orbit and swapping $\gamma$ and $\gamma'$ can be seen for example 
by following the orbits in Fig.\ \ref{fig3}.  As we later sum over all possible central periodic orbits, 
we fix $\gamma$ as having one traversal fewer than $\gamma'$ during its first encounter with the central 
periodic orbit and one traversal more during its second encounter. 

The two encounters of the orbit with the central periodic orbit have to be independent, because otherwise there exists no connected partner with a small but nonzero action difference. That means  the orbit has to decorrelate from the central periodic orbit. Then this orbit has to become ergodic before returning, therefore an encounter time, that is of the order of the Ehrenfest time is required, so the stretch away from the periodic orbit in Fig.~\ref{fig2,5} or the top loop in Fig.~\ref{fig2} must be of positive length. The total orbit  has thus to be longer than the sum of the two durations of the encounters with the central periodic orbit, $t_{{\rm enc,}1}+t_{{\rm enc,}2}$. This excludes the case that both $t_3$ and $t_5$ in Fig.~\ref{fig2,5} get so short that \textbf{both} stretches do not decorrelate from the central periodic orbit. One stretch is necessarily close to the 
periodic orbit, but when the other also becomes short and correlated with the central periodic orbit, 
it too must follow the periodic orbit closely.  The orbit $\gamma$ then only encounters the periodic orbit once, 
follows it for some number of traversals and then exits the system.  With no way to swap traversals between the 
different encounters, the partner $\gamma'$ is then identical to $\gamma$ and included in the diagonal approximation.

Orbital configurations with periodic orbit encounters as described above also occurred in the calculation of the covariance of the reflection coefficients \cite{Bro} yielding a term proportional to $\left(1-{\rm e}^{-2\tau_E/\tau_D} \right)$.  In fact, by cutting the top loop in Figure~\ref{fig2} and moving the ends to the correct places we can see we recreate Figure~\ref{fig1}.  Reversing this cutting though, to return to the transmission and reflection, we create the second periodic orbit which is the top loop in Fig.~\ref{fig2} and travels through $t_3$, the encounters and $t_4$ in Fig.~\ref{fig2,5}.  We will see that this changes the orbital configurations compared to the case of the variance, changing also the resulting contribution. For the covariance of the reflection coefficients it turned out to be essential \cite{Bro} to consider additionally to the encounter stretches, where both orbits in Fig.~\ref{fig1} are correlated with the central dashed-dotted periodic orbit, also the encounter fringes, where the orbits are correlated with themselves (or each other) \textbf{but} where they are no longer correlated to the periodic orbit.  We marked the places where correlations between fringes occurs in Fig.~\ref{fig1} by black vertical lines. The duration of the fringes before the orbits get correlated to the central periodic orbit is denoted by $t_s$ and after the orbits leave the central periodic orbit by $t_u$.

These fringes are the key to the difference between the possible orbital configurations for the covariance of the reflection coefficients on the one hand and the transmission and reflection on the other hand: In the case of the covariance of the reflection coefficients these fringes need to have a nonvanishing length, because the two orbits (see Fig.~\ref{fig1}) which are correlated during the fringes have to end at two different leads where they have to be uncorrelated.  The orbits away from the central periodic orbit must be long enough for the chaotic dynamics to allow this to happen.  
When we join one end of the dashed and one of the solid orbit in Fig.~\ref{fig1} say to return to the transmission (or reflection) as in Fig.~\ref{fig2} then it is no longer necessary that the upper periodic orbit (in Fig.~\ref{fig2}) thereby created has to be longer than the fringe times.  These fringes can now start to overlap as depicted in Fig.~\ref{fig3}; compared to Fig.~\ref{fig2,5} we 
let the fringes grow till they overlap in the link $t_3$ which itself is of positive duration.   The stretches of the orbit that connect to the leads must though still be longer than the duration of the fringes as they must decorrelate from the upper periodic orbit to exit the system.

Now we can explain the effect of the possible orbital configurations on the resulting contributions. For this we first review some details of calculations for obtaining contributions from orbits differing in encounters from \cite{Mul1,Heu1}. In order to count the number of orbits we use a sum rule based on the classical ergodicity which takes the form \cite{Ric}
\begin{equation}
\label{in6.5}
\sum _\gamma\left|B_\gamma\right|^2=N_1N_2\int_0^\infty p(t)
\end{equation}
with $N_1$ and $N_2$ the number of open transverse channels in the left and right lead respectively and $p(t)$ the survival probability of an orbit of duration time $t$. This probability decays exponentially for chaotic systems as $p(t)\sim{\rm e}^{-t/\tau_D}$ for $t\rightarrow\infty$ with $\tau_D=T_H/N$ and $N=N_1+N_2$.  
The encounters are characterized by the stable and unstable coordinate differences $s,u$ in a Poincar\'e surface of section inside each encounter between the central periodic orbit and the surrounding orbit. In terms of these coordinates the action difference between the two trajectories is given by \cite{Mul1,Bro}
\begin{equation}
\label{in7}
\Delta S=\sum_is_iu_i,
\end{equation}
where the sum runs over the different encounters of the considered orbit with the central periodic orbit. The length of each encounter is obtained by choosing that the considered orbit and the central periodic one close enough to each other that they can be linearized around each other \cite{Mul1}
\begin{equation}
\label{in8}
t_{{\rm enc,}i}(s_i,u_i)=\frac{1}{\lambda}\ln\left( \frac{c^2}{|s_iu_i|}\right)
\end{equation}
with $c$ of order one. During $t_{\rm enc}$, the survival probability is enhanced: either the orbit leaves the system during the first stretch or does not leave at all. A density of encounters $w_t(s,u)$ with respect to $s,u$ which characterizes the expected number of encounters the orbit has with the central periodic orbit can then be obtained, in the case here of two encounters with the periodic orbit, as \cite{Heu1}
\begin{equation}
\label{in9}
w_t(s,u)=\frac{1}{\Omega^2t_{{\rm enc},1}t_{{\rm enc},2}}\left( \prod_{i=1}^2\int dt_i\right)\int d\tau_p\int_0^{\tau_p}dt',
\end{equation}
where the phase space volume of the system under consideration is denoted by $\Omega$ and $t_i$ denotes the duration of two of the three links away from the periodic orbit (two connecting the opening to the central periodic orbit and one the periodic orbit to itself), $\tau_p$ is the duration of the central periodic orbit and $t'$ the time difference between the two points (in the different encounters) where each of the two encounter stretches reaches a phase space difference $c$ with respect to the periodic orbit.  The limits of the time integrals in (\ref{in9}) are determined by the fact that the duration of the links, the periodic orbit and the encounters have to be positive.  This differs from the treatment of Fig.~\ref{fig2,5} in \cite{Heu1} as they assumed that all five links have to have positive duration, but we allow some of them to overlap.  This automatically includes the other cases described at the start of this section as part of a continuous deformation of Fig.~\ref{fig2,5} or Fig.~\ref{fig2}.  In particular we allow $t_4$ and $t_5$ to shrink and instead just assume that $\tau_p$ is positive, so we therefore use this variable in (\ref{in9}).

Using these quantities, we obtain from the definition of the transmission $T$ (\ref{in6}) the following contribution resulting from the diagrams shown in Fig.~\ref{fig2} and Fig.~\ref{fig3}, which we denote $T^{\ref{fig2},\ref{fig3}}$ \cite{Heu1}
\begin{equation}
\label{in10}
T^{\ref{fig2},\ref{fig3}}\!=\left\langle \frac{N_1N_2}{T_H}\int_0^\infty dt\int_{-c}^cd^2{s}d^2{u}\,w_t(s,u){\rm e}^{\left(i/\hbar\right)\Delta S}p'(t)\right\rangle.
\end{equation}
We defined here the modified survival probability in the presence of encounters $p'(t)$ taking into account the modification mentioned after Eq.\ (\ref{in8}) for encounters within the fringes and with periodic orbits: when the encounters surround the periodic orbit the parts of the encounter stretches traversing a certain point of the periodic orbit are so close to each other that they either leave the cavity during the first traversal or do not leave at all \cite{Heu1,Bro} leading to
\begin{equation}
\label{in10.5}
p'(t)=p(t){\rm e}^{\left(t_{{\rm enc,}1}+t_{{\rm enc,}2}+t_s+t_u\right)/\tau_D}={\rm e}^{-\left(t_1+t_2+t_3+\tau_p\right)/\tau_D}. 
\end{equation}
This expression can be transformed, using (\ref{in9}) and converting the integral over the full duration of the orbit $t$ into one over the link $t_3$, into
\begin{eqnarray}
\label{eq1}
T^{\ref{fig2},\ref{fig3}} &=&\left\langle\frac{N_1N_2}{T_H}\left(\prod_{i=1}^3\int_0^\infty dt_i \exp\left( -\frac{t_i}{\tau_D}\right) \right)\int_{-c}^c d^2{s}d^2{u}\right. \nonumber\\&& \left.\times \int_0^\infty d\tau_p\int_{0}^{\tau_p} dt'\exp\left( -\frac{\tau_p}{\tau_D} \right)\frac{1}{\Omega^2 t_{{\rm enc},1}t_{{\rm enc},2}}\right.\nonumber\\&&\left. \times\exp\left(\frac{i}{\hbar}\sum_{i=1}^2{s_i}{u_i} \right)\right\rangle,
\end{eqnarray}
%
where we also used the explicit form of the survival probability and the action difference $\Delta S$. To understand that the expression in (\ref{eq1}) yields zero, we perform the integrals with respect to $s_i,u_i$, like in \cite{Bro} 
\begin{eqnarray}
\label{eq1.1}
&&\hspace*{-1.5em}\int_{-c}^cds_idu_i{\rm e}^{\left(i/\hbar\right)s_iu_i}\frac{1}{t_{\rm enc,i}}\nonumber\\&=&4c^2 \int_{0}^1dx_i\int_1^{1/x_i}d\sigma_i\cos\left(\frac{c^2x_i}{\hbar}\right) \frac{1}{\sigma_i t_{\rm enc,i}}\nonumber\\&=&4c^2\lambda\int_{0}^1dx_i\cos\left(\frac{c^2x_i}{\hbar}\right)
\end{eqnarray}
with the substitution $u_i=c/\sigma_i$ and $s_i=cx_i\sigma_i$. The integral in the last line in (\ref{eq1.1}) rapidly oscillates as a function of energy in the limit $\hbar\rightarrow 0$ and thus yields no contribution due to the energy average in ({\ref{eq1}).

We thus obtain that there are no quantum corrections (at least to this order) to the transmission when time reversal symmetry is absent
\begin{equation}
\label{eq1.2}
T^{\ref{fig2},\ref{fig3}}=0,
\end{equation}
and a similar calculation shows that this also holds for the reflection $R$. Coherent backscattering, i.e.\ having encounters at the opening that additionally have to be taken into account for reflection, is also not possible. First this
requires the encounter to be traversed in opposite direction on both
traversals, which can only occur with time reversal symmetry.  Second,
even with time reversal symmetry, when the trajectory returns to the
encounter the second time it would necessarily escape the systems, and
not be able to complete the rest of the semiclassical diagram.

To summarize, we saw in this section how, despite their close similarities, the two different orbital configurations appearing in the case of the covariance of the reflection on one hand and the transmission and reflection coefficients on the other lead to two different results: in the case of the covariance of the reflection to a term proportional to $\left(1-{\rm e}^{-2\tau_E/\tau_D} \right)$, in the case of the transmission and reflection coefficients to zero contribution.

\subsection{With time reversal symmetry}

We now turn to the calculations in the case with time reversal symmetry. In this case we also have to consider diagrams where the encounters are traversed in different directions by the orbit. As their contributions are quite different we will study them individually. We start with two independent encounters with no central periodic orbit involved, referred to as two 2-encounters, shown in Fig.~\ref{fig4}.
\begin{figure}
\begin{center}
\includegraphics[width=8cm]{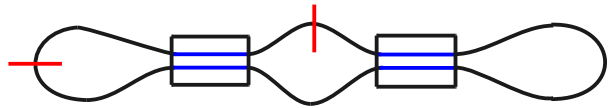}
\caption{Periodic orbit with two independent 2-encounters. The different positions, where it can be cut to obtain an open orbit contributing to the transmission are indicated by (red) perpendicular lines, the position of the (blue) encounter stretches are indicated by a box.}
\label{fig4}
\end{center}
\end{figure}
We first cut the periodic orbit during one of the middle links and refer to the corresponding contribution as $T^{\ref{fig4}a}$. In this case the contribution of the two $s,u$-integrals for the two different encounters factorizes and can be evaluated for each encounter separately, as was done in \cite{Bro} for the case of the reflection covariance (obtained by cutting both the leftmost and rightmost links in the periodic orbit in Fig.~\ref{fig4}). Each encounter provides a factor $-N{\rm e}^{-\tau_E/\tau_D}$, the five links factors $N^{-1}$ and the leads the factor $N_1N_2$ so that we obtain for the contribution $T^{\ref{fig4}a}$,
\begin{equation}
\label{eq2}
T^{\ref{fig4}a}=\frac{N_1N_2}{\left(N_1+N_2 \right)^3} {\rm e}^{-2\tau_E/\tau_D}.
\end{equation}
The corresponding contribution to the reflection $R^{\ref{fig4}a}$ is obtained by multiplying $T^{\ref{fig4}a}$ by $N_1/N_2$ to take into account that the orbit leaves through the lead 1 instead of lead 2. When cutting the left link of the periodic orbit in Fig.~\ref{fig4}, whose contribution we denote $T^{\ref{fig4}b}$, we obtain for the transmission the same result as $T^{\ref{fig4}a}$. However for the reflection in this case it is also possible to obtain a coherent backscattering contribution by shrinking the length of both links on the left in Fig.~\ref{fig4} to zero (or we cut the diagram in Fig.~\ref{fig4} at the leftmost encounter and move this to the lead).  Also in this case the encounter integrals for the two encounters factorize, yielding
\begin{equation}
\label{eq3}
R^{\ref{fig4}b}=\frac{N_1^2}{\left(N_1+N_2 \right)^3} {\rm e}^{-2\tau_E/\tau_D}-\frac{N_1}{\left(N_1+N_2 \right)^2} {\rm e}^{-2\tau_E/\tau_D},
\end{equation}
where the first term is the same as $R^{\ref{fig4}a}$ and the second comes from the coherent backscattering.

Next we consider, as for the case of no time reversal symmetry, the situation of two 2-encounters near a periodic orbit. The configuration where the encounter stretches are parallel (in the same direction) was treated in the last subsection so, as we have now the freedom to traverse the two encounter stretches in opposite directions, we now turn to configurations where some of the stretches are antiparallel to each other.  Starting with the periodic orbit configuration in Fig.~\ref{fig5,5} (a 3-encounter in \cite{Mul1}) there are three possible places to cut this orbit open as shown by the red lines perpendicular to the orbit. By opening the parts not enclosing the central periodic orbit, we obtain a configuration shown in Fig.~\ref{fig6}.
\begin{figure}
\begin{center}
\includegraphics[width=8cm]{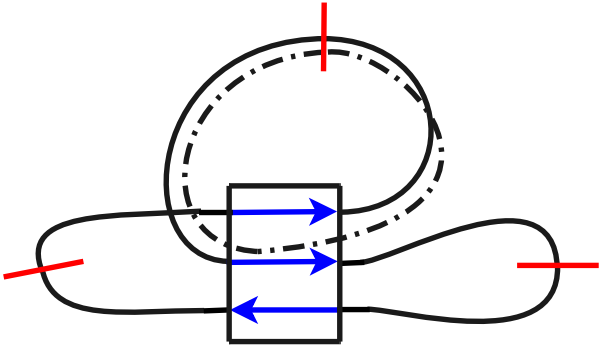}
\caption{A periodic orbit encounter only existing in the case of time reversal symmetry. The central periodic orbit is drawn with a  dashed-dotted line, the position of the (blue) encounter stretches are marked by a box. The (red) lines perpendicular to the orbit mark the places where it can be cut open.}
\label{fig5,5}
\end{center}
\end{figure}
\begin{figure}
\begin{center}
\includegraphics[width=8cm]{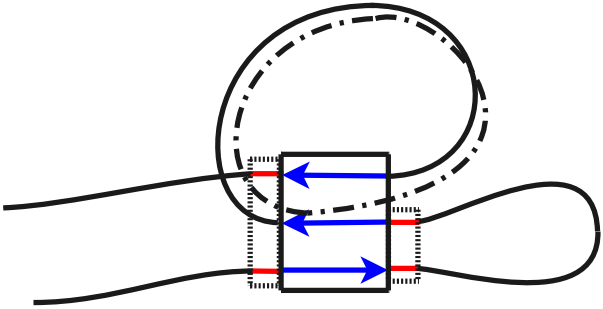}
\caption{A periodic orbit encounter only existing in the case of time reversal symmetry. The position of the (blue) encounter stretches are marked by a box, the position of the fringes by small dotted boxes.}
\label{fig6}
\end{center}
\end{figure}
%

Unlike the case without time reversal symmetry, we can see that some of the fringes must have non-vanishing length like considered in \cite{Bro}: the two fringes marked by dotted boxes in Fig.~\ref{fig6} on the right hand side of the encounter cannot have vanishing length, because as long as the two parts are correlated, the corresponding loop they form cannot close. The two fringes in Fig.~\ref{fig6} on the left hand side of the encounter can only vanish in the case of coherent backscattering, i.e.\ if the orbit starts and ends in a correlated manner in the same lead.  Note that in the left fringe (defined where stretches are correlated with each other away from the central periodic orbit) we only have the two stretches which connect to the leads and that the remaining encounter stretch in Fig.~\ref{fig6} which follows the central periodic orbit has already decorrelated from the others so it does not also need to escape. To evaluate the contribution we first need to determine the values of the prefactors $a,b,d$ in the exponential in $J$ in the Appendix in front of $t_{{\rm enc},1}+t_{{\rm enc},2}$, $t_s+t_u$ and $\tau_p$, respectively. As the survival probability along the periodic orbit depends only on $\tau_p$ and not on $t_{{\rm enc},1},t_{{\rm enc},2}$, we obtain $a=0$ and $d=1/\tau_D$. During the fringes we have two correlated stretches with the survival probability determined by one of them, thus yielding $b=-1/\tau_D$. When multiplying the resulting contribution for $J$ by the factors resulting from the links and the channel factors due to the leads 
we obtain the contribution $T^{\ref{fig6}}$ originating from Fig.~\ref{fig6} to the transmission 
\begin{equation}
\label{eq4}
T^{\ref{fig6}}=\frac{N_1N_2}{2\left( N_1+N_2\right)^3 }\left(1-{\rm e}^{-2\tau_E/\tau_D} \right)
\end{equation}
and to the reflection
\begin{eqnarray}
\label{eq5}
R^{\ref{fig6}}&=&\frac{N_1^2}{2\left( N_1+N_2\right)^3 }\left(1-{\rm e}^{-2\tau_E/\tau_D} \right)\nonumber\\&-&\frac{N_1}{2\left( N_1+N_2\right)^2}\left(1-{\rm e}^{-2\tau_E/\tau_D} \right).
\end{eqnarray}
%

The last case, depicted in Fig.~\ref{fig7}, is obtained by opening along the central periodic orbit in Fig.~\ref{fig5,5}.
\begin{figure}
\begin{center}
\includegraphics[width=8cm]{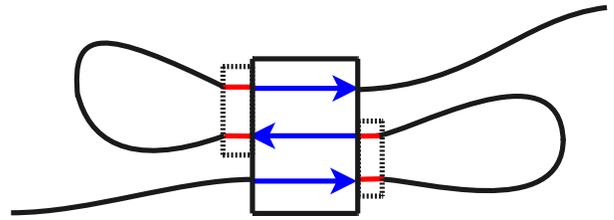}
\caption{A 3-encounter involving no periodic orbits. The (blue) encounter stretches are again marked by boxes, the places where fringe correlations can occur (are marked red) indicated by smaller dotted boxes.}
\label{fig7}
\end{center}
\end{figure}
In general any two of the three stretches on either side of the encounter could remain correlated in the fringes away from the main encounter where all three stretches are close and correlated. The duration of the fringes, i.e.\ here in general the orbital parts where only two of the three encounter stretches are correlated, is denoted before and after where all three orbits are correlated by $t_s$ and $t_u$, respectively as in \cite{Bro1} and in  the Appendix. On each side, fringe correlations become important if the two orbital parts containing the fringes are connected to each other, referred to as case A, but not if one orbital part of them is connected to the opening, referred to as case B. The reason why we have to take into account fringe correlations in case A is that namely the loop cannot close as long as the two parts of the orbit are still correlated.  In case B the part of the orbit connected to the opening still has to be longer than the fringes  so that when it escapes it does not force the rest of the orbit to also escape.  However the other fringe which lies on the orbit that is not connected to the opening in Fig.~\ref{fig7} has no length restriction: if the length of that fringe tends to zero, the orbital part connected to the opening will just follow the first one for the time $t_s$ or $t_u$. The latter part then also contains the survival probability contribution due to the fringes. The fact, that \textit{only} the part of the orbit connected to the opening has a length restriction due to the fringes together with the enhancement of the survival probability during the fringe parts lets, as already in (\ref{eq1}), the $t_s$ and the $t_u$ drop from the resulting expressions for the contribution $T^{\ref{fig7}}$ from these diagrams in case B.

With these remarks in mind we evaluate the contribution $T^{\ref{fig7}}$ in Fig.~\ref{fig7} by making use of the results obtained in \cite{Bro1}, that we review in the Appendix as contributions $K_1$ and $K_2$. 
The overall contribution $K$ is split into two parts $K_1$ and $K_2$: $K_1$ contains the contribution resulting from the 3-encounter without fringes and $K_2$ the contribution resulting from the difference between the 3-encounter with fringes and a 3-encounter without fringes. In all the cases considered here we include the first part $K_1$ where as for the survival probability during encounters we only need to count one encounter stretch we get $f=-1/\tau_D$ in $K_1$ in the Appendix. To obtain the contribution $K_2$ in this case we first note that it was shown in \cite{Bro1} to be sufficient to only consider certain encounter diagrams: only one stretch contains two fringes, the other two fringes lie on the other stretches in a certain way. Furthermore by setting $g_1=0$ or $g_2=0$, implying that only $t_s$ or $t_u$ is nonzero we obtain zero contribution, see Eq.~(\ref{ap7}). Thus three different possibilities remain: one from each of the three stretches containing two fringes, one belonging to case A and two to case B. As already explained there is no $t_s$, $t_u$-dependence in case B and thus in this case no contribution to $K_2$. In case A we obtain $1/3$ of the contribution $K_2$ in Eq.~(\ref{ap6}), we set $f=g=-1/\tau_D$ to take into account that only one stretch of the encounter is taken into account in the survival probability.  
%
%
%
%
%
%
For the overall contribution $T^{\ref{fig7}}$ we therefore have
\begin{equation}
\label{eq9}
T^{\ref{fig7}}=-\frac{N_1N_2}{\left(N_1+N_2 \right)^3 }{\rm e}^{-2\tau_E/\tau_D}.
\end{equation}
The same contribution times the factor $N_1/N_2$ is obtained for the reflection.

To calculate the overall quantum correction to the transmission at the considered order $T_{\rm 2nd}$ in the case of time reversal symmetry we sum twice (to account for diagrams related by symmetry) the contributions from (\ref{eq2}) and (\ref{eq4}) and the contribution from (\ref{eq9}) yielding
\begin{equation}
\label{eq9.1}
T_{\rm 2nd}=\frac{N_1N_2}{\left( N_1+N_2\right)^3 }.
\end{equation}
Note that this quantum correction is independent of the Ehrenfest time. This also holds for the corresponding contribution to the reflection $R_{\rm 2nd}$, which we obtain here by adding the contribution from (\ref{eq3}) to twice the contribution from (\ref{eq5}) and to the related contributions from (\ref{eq2}) and (\ref{eq9}) multiplied by $N_1/N_2$ 
\begin{equation}
\label{eq9.2}
R_{\rm 2nd}=\frac{N_1^2}{\left( N_1+N_2\right)^3}-\frac{N_1}{\left( N_1+N_2\right)^2}.
\end{equation}
\subsection{Current conservation}
Having calculated all contributions to the transmission and reflection we now want to check if current conservation is fulfilled, i.e.\ if the transmission and the reflection calculated for one lead add up to the number of open channels in that lead.  As without time reversal symmetry there are no contributions at the order $1/N$ considered here, current conservation, already fulfilled at the diagonal level, is thus not violated. In the case of time reversal symmetry 
%
%
the contributions to $T$ and $R$ at the considered order are given in (\ref{eq9.1}) and (\ref{eq9.2}) and sum to zero. Current conservation is thus again fulfilled.  We want to emphasize here that correlations between encounter fringes, first treated in \cite{Bro}, were important to obtain this result: forgetting for a moment the effect of fringe correlations, the contribution (\ref{eq9}) would possess  the Ehrenfest-time dependence ${\rm e}^{-\tau_E/\tau_D}$ and the contributions (\ref{eq4},\ref{eq5}) would be zero leading to a non current conserving result for $T_{\rm 2nd}$ and $R_{\rm 2nd}$. 


\section{Frequency dependence}
In this section we want to generalize the results obtained for dc-transport to the ac-case \cite{Petit}, i.e.\ we want to consider
\begin{equation}
\label{eq10.1}
T(\omega)=\left\langle {\rm Tr}(t(E+\hbar\omega/2)t^\dagger(E-\hbar\omega/2))\right\rangle
\end{equation}
and a correspondingly defined $R(\omega)$.
As the calculation leading to this generalization is straightforward we only briefly explain the difference to the calculation before and then show the results. In general adding a frequency dependence means including into the formulas in section II a factor ${\rm e}^{i\omega t}$ with the overall duration $t$ of the orbit. 
In the case of no time reversal symmetry we get in terms of the notation of eq.\ (\ref{eq1}) an additional factor ${\rm e}^{i\omega\left(\tau_p+t_{enc,1}+t_{enc,2}\right)}\prod_{i=1}^3{\rm e}^{i\omega t_i}$. To include this factor when performing the $s,u$-integrals we take $a$, $b$, $d$ for $\omega=0$ from the last section and include the $\omega$-dependent exponential factor given in the last sentence  to obtain $a=i\omega$, $b=0$ and $d=1/\tau_D-i\omega$. Inserting this in $J$ in the Appendix and taking into account the factors from the links and the leads we obtain
\begin{equation}
\label{eq10.2}
T^{\ref{fig2},\ref{fig3}}(\omega)=\frac{N_2}{N_1}R^{\ref{fig2},\ref{fig3}}(\omega)=\frac{-N_1N_2}{\left(N_1+N_2 \right)^3} \frac{\left(\omega\tau_D \right)^2 }{\left(1-i\omega\tau_D \right)^5}{\rm e}^{2i\omega\tau_E}.
\end{equation}

In the orthogonal case, including a frequency dependence into eq.\ (\ref{eq2}) adds an additional factor ${\rm e}^{i\omega\left(2t_{enc,1}+2t_{enc,2}\right)}\prod_{i=1}^5{\rm e}^{i\omega t_i}$ yielding finally 
\begin{eqnarray}
\label{eq11}
T^{\ref{fig4}a}(\omega)&=&T^{\ref{fig4}b}(\omega)=\frac{N_2}{N_1}R^{\ref{fig4}a}(\omega)\nonumber\\&=&\frac{N_1N_2}{\left(N_1+N_2 \right)^3} \frac{\left(1-2i\omega\tau_D \right)^2 }{\left(1-i\omega\tau_D \right)^5}{\rm e}^{-2\tau_E/\tau_D+4i\omega\tau_E}.\nonumber\\
\end{eqnarray}
The first term of eq.\ (\ref{eq3}) is modified in the same way as the expression in eq.\ (\ref{eq11}) while for the second we have three links instead of five, reducing the power of $\left(1-i\omega\tau_D \right)$ in the denominator by two, and an additional integral over the duration of the encounter reducing the power of $\left(1-2i\omega\tau_D \right)$ by one compared to the first term. We thus obtain
\begin{eqnarray}
\label{eq12}
R^{\ref{fig4}b}(\omega)&=&\frac{N_1^2}{\left(N_1+N_2 \right)^3} \frac{\left(1-2i\omega\tau_D \right)^2 }{\left(1-i\omega\tau_D \right)^5}{\rm e}^{-2\tau_E/\tau_D+4i\omega\tau_E}\nonumber\\&-&\frac{N_1}{\left(N_1+N_2 \right)^2}\frac{\left(1-2i\omega\tau_D \right)}{\left(1-i\omega\tau_D \right)^3} {\rm e}^{-2\tau_E/\tau_D+4i\omega\tau_E}.\nonumber\\
\end{eqnarray}
In equation (\ref{eq4}) the additional factor ${\rm e}^{i\omega\left( \tau_p+t_{enc,1}+t_{enc,2}+2t_s+2t_u\right)}\prod_{i=1}^3{\rm e}^{i\omega t_i}$ occurs, the equation is thus replaced by
\begin{eqnarray}
\label{eq14}
T^{\ref{fig6}}(\omega)&=&\frac{N_1N_2}{2\left( N_1+N_2\right)^3 }\left[\left( {\rm e}^{2i\omega\tau_E}-{\rm e}^{-2\tau_E/\tau_D+4i\omega\tau_E} \right)\right. \nonumber \\ &\times&\left. \frac{\left(1-2i\omega\tau_D \right)^2}{\left(1-i\omega\tau_D \right)^5}-\frac{2\omega^2\tau_D^2}{\left(1-i\omega\tau_D \right)^5}{\rm e}^{2i\omega\tau_E}\right].
\end{eqnarray}
The latter equation can be obtained from $J$ in the Appendix by setting $a=i\omega$, $b=-1/\tau_D+2i\omega$ and $d=1/\tau_D-i\omega$, again considering the additional terms from the $\omega$-dependent exponentials.
The additional frequency in the first term in eq.\ (\ref{eq5}) has the same effect as in eq.\ (\ref{eq4}), in the second term we again have one instead of three link times $t_i$ and an additional integral over $t_s$ or $t_u$
\begin{eqnarray}
\label{eq15}
R^{\ref{fig6}}(\omega)&=&\frac{N_1^2}{2\left( N_1+N_2\right)^3 }\left[\left( {\rm e}^{2i\omega\tau_E}-{\rm e}^{-2\tau_E/\tau_D+4i\omega\tau_E} \right)\right.\nonumber \\  &\times& \left. \frac{\left(1-2i\omega\tau_D \right)^2}{\left(1-i\omega\tau_D \right)^5}-\frac{2\omega^2\tau_D^2}{\left(1-i\omega\tau_D \right)^5}{\rm e}^{2i\omega\tau_E}\right]\nonumber\\&-&\frac{N_1}{2\left( N_1+N_2\right)^2}\left[\left( {\rm e}^{2i\omega\tau_E}-{\rm e}^{-2\tau_E/\tau_D+4i\omega\tau_E}\right)\right. \nonumber\\&\times&\left. \frac{\left(1-2i\omega\tau_D \right)}{\left(1-i\omega\tau_D \right)^3} \right].
\end{eqnarray}
In case of eq.\ (\ref{eq9}) we get by taking the corresponding contribution of the encounter again from the Appendix with $f=-1/\tau_D+3i\omega$ and $g=-1/\tau_D+2i\omega$ in $K_1$ and $K_2$ since $t_{\rm enc}$ is traversed three times and the fringes two times  
\begin{eqnarray}
\label{eq16}
T^{\ref{fig7}}(\omega)&=&-\frac{N_1N_2}{\left(N_1+N_2 \right)^3 }\left[ \frac{\left(1-2i\omega\tau_D \right)^2}{\left(1-i\omega\tau_D \right)^5}{\rm e}^{-2\tau_E/\tau_D+4i\omega\tau_E}\right.\nonumber\\ &&\left.+\frac{\omega^2\tau_D^2}{\left(1-i\omega\tau_D \right)^5}{\rm e}^{-\tau_E/\tau_D+3i\omega\tau_E}\right] 
\end{eqnarray}
and a corresponding contribution for the reflection.
%
%
%

After obtaining these results it is now possible to check if they fulfill the relation
\begin{equation}
\label{eq18}
\frac{d}{d\tau_E}\left.\frac{d}{d\omega}{\rm Tr}\left[S\left(E+\hbar\omega \right) S^{\dagger}\left(E-\hbar\omega \right)\right] \right|_{\omega=0}=0
\end{equation}
with the scattering matrix at the energy $E$, $S(E)$ containing the reflection and transmission subblocks $r, t$ for the incoming wave in the lead 1 and $r',t'$ for the incoming wave in the lead 2, respectively
\begin{equation}
\label{eq18.01}
S(E)=\left(\begin{matrix}r(E)&t'(E)\\t(E)&r'(E)\end{matrix}\right).
\end{equation}
Before we only considered the reflection and transmission for an incoming wave in the lead 1, i.e.\ only the correlators of elements of $r(E)$ and $t(E)$. The corresponding results for the correlators of $r'(E)$ and $t'(E)$ are obtained by swapping $N_1$ and $N_2$. 

In order to see why relation (\ref{eq18}) is fulfilled we rewrite it in terms of the Wigner time delay \cite{Wig}, measuring the additional time spend in the scattering process compared to the free motion, $\tau_W\equiv\left.\frac{d}{d\omega}{\rm Tr}\left[S\left(E+\hbar\omega \right) S\left(E-\hbar\omega \right) \right] \right|_{\omega=0}$. Equation (\ref{eq18}) is then
\begin{equation}
\label{eq18.1}
\frac{d}{d\tau_E}\tau_W=0.
\end{equation}
That this relation has to hold can be obtained by comparing the two equivalent representations of the Wigner time delay discussed in \cite{Wig}; their semiclassical equivalence is discussed in \cite{Kui}. The first representation in terms of the density of states involves a single sum over trapped periodic orbits, the second representation in terms of transmission coefficients involves a double sum over lead-connecting paths. As the first representation yields, after taking an energy average, an Ehrenfest time independent result - we cannot identify any Ehrenfest time dependent contributions in a single sum over periodic orbits - $\tau_W$ has to be Ehrenfest time independent.

In terms of the subblocks of $S(E)$ introduced in eq.\ (\ref{eq18.01}), $\tau_W$ can be expressed as
\begin{eqnarray}
\label{eq18.11}
\tau_W&=&\frac{d}{d\omega}\left|\left[T(\omega)+R(\omega)+T'(\omega)+R'(\omega)\right]\right|_{\omega=0} 
\end{eqnarray}
with the primes again denoting that the incoming wave is in lead 2 instead of lead 1.
We start our further analysis of the first two terms: 
In order to check if our results for $T(\omega)$ and $R(\omega)$ given above eq.\ (\ref{eq18}) fulfill relation (\ref{eq18.1}), we first consider the sum of the contributions to $\tau_W$ which decrease with increasing Ehrenfest time. The contribution proportional to ${\rm e}^{-\tau_E/\tau_D}$ is obtained by considering the corresponding term in (\ref{eq16}) yielding 
\begin{equation}
\label{eq19}
\left.\frac{d}{d\omega}\left[-\frac{N_1}{\left( N_1+N_2\right)^2}\frac{\omega^2\tau_D^2}{\left(1-i\omega\tau_D \right)^5}{\rm e}^{-\tau_E/\tau_D+3i\omega\tau_E}\right]\right|_{\omega=0}.
\end{equation}
For calculating the contribution proportional to ${\rm e}^{-2\tau_E/\tau_D}$ we sum the corresponding terms from (\ref{eq11}), (\ref{eq12}) and (\ref{eq16}) 
\begin{equation}
\label{eq20}
\left.\frac{d}{d\omega}\left[\frac{N_1\omega^2\tau_D^2}{\left( N_1+N_2\right)^2}\frac{\left(1-2i\omega\tau_D \right)}{\left(1-i\omega\tau_D \right)^5}{\rm e}^{-2\tau_E/\tau_D+4i\omega\tau_E}\right] \right|_{\omega=0}.
\end{equation}
In the case of the contributions increasing or oscillating with increasing Ehrenfest time we obtain from (\ref{eq14}) and (\ref{eq15})
\begin{eqnarray}
\label{eq21}
\left.\frac{d}{d\omega}\left\lbrace \frac{-N_1}{\left( N_1+N_2\right)^2}\frac{\omega^2\tau_D^2}{\left(1-i\omega\tau_D \right)^5}\left[2{\rm e}^{2i\omega\tau_E}\right.\right.\right.\nonumber\\ \left.\left.\left.-\left(1-2i\omega\tau_D\right) \left({\rm e}^{2i\omega\tau_E}-{\rm e}^{-2\tau_E/\tau_D+4i\omega\tau_E}\right)  \right] \right\rbrace \right|_{\omega=0}
\end{eqnarray}
and from (\ref{eq10.2})
\begin{equation}
\label{eq22}
\left.\frac{d}{d\omega}\left[ -\frac{N_1}{\left(N_1+N_2 \right)^2} \frac{\left(\omega\tau_D \right)^2 }{\left(1-i\omega\tau_D \right)^5}{\rm e}^{2i\omega\tau_E} \right] \right|_{\omega=0},
\end{equation}
which is the only contribution also existing in the absence of time reversal symmetry.

The results in (\ref{eq19}-\ref{eq22}) fulfill eq.\ (\ref{eq18}), because all are proportional to $\omega^2$ and thus are equal to zero after differentiating them with respect to $\omega$ and setting $\omega=0$.

The results obtained from the two second terms in eq.\ (\ref{eq18.11}) differ from the first ones by a factor $N_2/N_1$ and thus also yield zero contribution to $\tau_W$.

\section{Spectral form factor}

In this section we want to apply our knowledge about the orbital configurations which contribute to the conductance to calculate the first off-diagonal quantum correction to the spectral form factor.  We first want to briefly review the contributions calculated in \cite{Rah} so we will use almost the same notation as there and for further details we refer the reader to that paper. For systems with time reversal symmetry the first correction derives from the orbit pair depicted in Fig.~\ref{fig0} whose Ehrenfest time dependence is simply ${\rm e}^{-\tau_E/\tau_D}$.  For systems without time reversal symmetry however we have diagrams starting like in Fig.~\ref{fig2,5} but with the orbits in the leads connected together so that $t_1$ and $t_2$ join to a single link.  Note that we can then identify four periodic orbits in the picture, one central orbit, one through $t_3$ as before and two through the newly joined links and $t_3$ and $t_5$, respectively.  From there we can allow the encounters to overlap to create a 3-encounter and then finally to wind around the central periodic orbit, as described at the start of section~\ref{unitrans}.  As we saw for the transmission there are further possibilities compared to the covariance of the reflection (or we can relax more restrictions) and likewise here there are additional contributions.   We will see how they lead semiclassically to the field-theoretical result for the first off-diagonal quantum correction to the spectral form factor, but first we recall the results of the diagrams covered in \cite{Rah}.

In \cite{Rah} the contribution of two independent 2-encounters (c.f.\ Fig.~\ref{fig2,5}), denoted by $\delta K_{2b}(\omega) $, is given by 
\begin{equation}
\label{ff1}
\delta K_{2b}(\omega)=\frac{1}{2\pi \hbar T_H^2}\frac{\partial^2}{\partial\omega^2}\frac{\rm{e}^{4i\omega\tau_E}}{\omega^2},
\end{equation}
the contribution of one 3-encounter, $\delta K_{2c}(\omega)$, obtained by allowing the encounter stretches to overlap along one enclosed periodic orbit in Fig.~\ref{fig2,5} at one end, is obtained using \cite{Bro1}  
to be 
\begin{equation}
\label{ff2}
\delta K_{2c}(\omega)=\frac{1}{2\pi\hbar T_H^2}\frac{\partial^2}{\partial\omega^2}\frac{1}{\omega^2}\left(3\rm{e}^{3i\omega\tau_E}-4\rm{e}^{4i\omega\tau_E} \right).
\end{equation}
A further diagram results from encounter overlap along one enclosed periodic orbit at both ends, see Fig.~\ref{fig8}. The overall contribution $I$ containing the contribution from the latter diagram and the contributions $\delta K_{2b}(\omega)$ and $\delta K_{2c}(\omega)$ is obtained by considering $J$ in the appendix with $a=i\omega$, $b=2i\omega$ and $d=-i\omega$, because the orbit is assumed to be longer than $t_{{\rm enc},1}+t_{{\rm enc},2}+2t_s+2t_u$, and multiplying it by factors resulting from the links not surrounding the central periodic orbit. A technical complication is that this diagram contains three copies of the contribution with a 3-encounter and 4 copies of the contribution with two 2-encounters.  Naturally we only want to include one copy later so we subtract them all here.  All told, the contribution resulting from the periodic orbit encounters, $\delta K_{2d}\left(\omega \right)$ was calculated in \cite{Rah} to be 
\begin{eqnarray}
\label{ff3}
\delta K_{2d}(\omega)&=& I-4\delta K_{2b}(\omega)-3\delta K_{2c}(\omega)\nonumber\\&=&\frac{1}{2\pi\hbar T_H^2}\frac{\partial^2}{\partial\omega^2}\frac{1}{\omega^2}\left(3\rm{e}^{2i\omega\tau_E}-9\rm{e}^{3i\omega\tau_E}\right. \nonumber\\ &&\left.+6\rm{e}^{4i\omega\tau_E}\right).
\end{eqnarray} 

\begin{figure}
\begin{center}
\includegraphics[width=7cm]{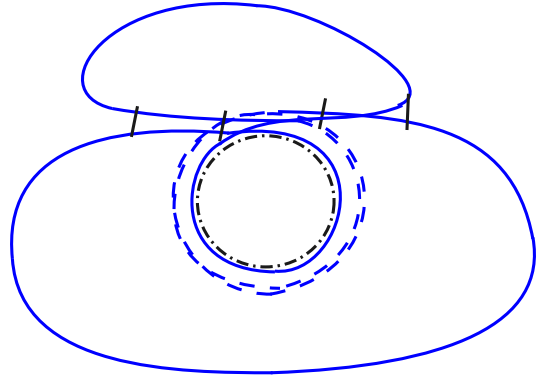}
\caption{A  diagram accounted for in the contribution $\delta K_{2d}(\omega)$ to the spectral form factor. A central dashed-dotted periodic orbit is encountered two times. Fringe correlations are marked by black vertical lines. For the partner (not shown) one traversal of the central periodic orbit is exchanged between the first and the second encounter.}
\label{fig8}
\end{center} 
\end{figure}

However, as already explained above, this is only the complete set of contributions in the case when the two orbits approaching and leaving the periodic orbit are open like in the case of the reflection covariance as in Fig.~\ref{fig1}, otherwise the corresponding orbital parts can get shorter than the duration of the fringes, like moving from Fig.~\ref{fig2} to Fig.~\ref{fig3}. To take into account this additional configuration we replace $\delta K_{2d}(\omega)$ by another contribution denoted by $\delta K_{2e}\left(\omega\right)$. 
Up to now we only allowed the orbital parts decorrelated from the central periodic orbit, to be longer than $2t_s+2t_u$.  However the contribution considered now results from an orbital configuration where the two other links, those decorrelated from the central periodic orbit in Fig.~\ref{fig8}, get shorter than $t_s+t_u$ each. The bottom and the top loop in Fig.~\ref{fig8} outside of the encounter with the central periodic orbit must again have positive length but do not necessarily need to be longer than the fringes. 
In order to calculate this contribution we first consider $J$ in the Appendix with $a=i\omega$, $b=0$ and $d=-i\omega$, because the orbit has a minimal length of $t_{{\rm enc,}1}+t_{{\rm enc,}2}$ as in Eq.~(\ref{eq10.2}), together with the factors resulting from the links not surrounding the central periodic orbit and subtract from this contribution like in Eq.~(\ref{ff3}) $\delta K_{2b}(\omega)$ and $\delta K_{2c}(\omega)$ with the right multiplicity factors. The corresponding contribution denoted by $\delta {K}_{2e,1}(\omega)$ is 
\begin{eqnarray}
\label{ff3.5}
\delta {K}_{2e,1}\!(\omega)&\!\!=\!\!&\frac{1}{2\pi\hbar T_H^2}\frac{\partial^2}{\partial\omega^2}\frac{1}{\omega^2} \rm{e}^{2i\omega\tau_E}\!\!\!\!-4\delta K_{2b}(\omega)-3\delta K_{2c}(\omega)\nonumber\\&\!\!=\!\!&\frac{1}{2\pi\hbar T_H^2}\frac{\partial^2}{\partial\omega^2}\frac{1}{\omega^2}\left[\rm{e}^{2i\omega\tau_E}-9\rm{e}^{3i\omega\tau_E}+8\rm{e}^{4i\omega\tau_E}\right].\nonumber\\ 
\end{eqnarray}
This procedure however counts some configurations containing a surrounded periodic orbit twice: shrinking in Fig.~\ref{fig8} the length of the upper periodic orbit to zero we again obtain a contribution containing one surrounded periodic orbit. A configuration containing one surrounded periodic orbit was however already taken into account in $\delta K_{2e,1}(\omega)$ when shrinking the length of the central periodic orbit to zero. We thus subtract the latter contribution. 
This contribution is calculated by again making use of $K_1$ and $K_2$ in the Appendix: we therefore consider a 3-encounter, i.e.\ $f=3i\omega$, with fringes with duration between $t_s+t_u$ and $2t_s+2t_u$. We thus consider once the prefactor $g=2i\omega$ and once $g=i\omega$ in front of $t_s+t_u$ and take the difference of the two results obtained for $K$ in the Appendix yielding the contribution $\delta {K}_{2e,2}(\omega)$ given by 
\begin{eqnarray}
\label{ff3.6}
\delta {K}_{2e,2}(\omega)&=&\frac{1}{2\pi\hbar T_H^2}\frac{\partial^2}{\partial\omega^2}\frac{1}{\omega^2}\left[{\rm e}^{3i\omega\tau_E}-{\rm e}^{2i\omega\tau_E}\right.\nonumber\\ &-&\left.4\left({\rm e}^{4i\omega\tau_E}-{\rm e}^{3i\omega\tau_E}\right)\right].
\end{eqnarray}
Adding the two contributions to $\delta K_{2e}(\omega)$ we obtain 
 \begin{eqnarray}
\label{ff4}
\delta K_{2e}(\omega)&=&\delta {K}_{2e,1}(\omega)+\delta {K}_{2e,2}(\omega)\nonumber\\&=&\frac{1}{2\pi\hbar T_H^2}\frac{\partial^2}{\partial\omega^2}\frac{4}{\omega^2}\left(\rm{e}^{4i\omega\tau_E}-\rm{e}^{3i\omega\tau_E}\right). 
\end{eqnarray} 

Summing now the quantum corrections in the absence of time reversal symmetry given in (\ref{ff1}), (\ref{ff2}) and (\ref{ff4}) we obtain for the overall quantum correction at the considered order $\delta K(\omega)$
\begin{equation}
\label{ff5}
\delta K(\omega)=\frac{1}{2\pi \hbar T_H^2}\frac{\partial^2}{\partial\omega^2}\frac{\left( \rm{e}^{4i\omega\tau_E}-\rm{e}^{3i\omega\tau_E}\right)}{\omega^2}.
\end{equation}
This yields then after the Fourier transform for the corresponding correction to the spectral form factor $\delta K(t)$
\begin{equation}
\label{ff6}
\delta K(\tau)=-\frac{\tau^2}{2\pi\hbar}\left[\Theta\left(\tau T_H-3\tau_E\right)-\Theta\left(\tau T_H-4\tau_E\right) \right] 
\end{equation}
with $\Theta(x)\equiv\int_0^xdx'\theta(x')=x\theta(x)$ with the Heaviside theta function $\theta(x)$. Expression (\ref{ff6}) was also obtained in \cite{Tia} by field-theoretical methods. 

As already noted for the conductance, we also want to emphasize here that these results for the spectral autocorrelation function could not be obtained without considering fringes: not doing so we would only get the contribution (\ref{ff1}) with a multiplicity factor four along with the contribution from a 3-encounter with 3 equally long encounter stretches, given by 
\begin{equation}
\label{ff7}
-\frac{1}{2\pi\hbar T_H^2}\frac{\partial^2}{\partial\omega^2}\frac{{\rm e}^{3i\omega\tau_E}}{\omega^2}
\end{equation}
with a multiplicity factor three and the overall contribution resulting from all possible encounter configurations given by
\begin{equation} 
\label{ff8}
\frac{1}{2\pi\hbar T_H^2}\frac{\partial^2}{\partial\omega^2}\frac{{\rm e}^{2i\omega\tau_E}}{\omega^2}
\end{equation}
with a multiplicity factor one. As one can easily see it is not possible to obtain the field-theoretical result from just these semiclassical contributions.

\section {Conclusions}
In this paper we have shown how to calculate the ($1/N$) quantum correction to the transmission and reflection for systems both with and without time reversal symmetry.  Starting with dc-transport, we obtained at the considered order that the transmission as well as the reflection are zero in the case of no time reversal symmetry.  In the presence of time reversal symmetry the overall contributions to the transmission (\ref{eq9.1}) and the reflection (\ref{eq9.2}) are independent of the Ehrenfest time and fulfill current conservation.  This simply means that the quantum corrections to the transmission and the reflection add up to zero. We extended this analysis then to the ac-transport by including a finite energy difference $\hbar\omega$ between the two scattering matrix elements. For the Wigner time delay we saw that the results led to zero extra contribution and importantly that there is no Ehrenfest time dependence consistent with the two complementary semiclassical representations of the time delay.

For the transmission and reflection the key step is that we can relax one of the restrictions compared to the calculation of the reflection covariance in \cite{Bro}.  Namely, because of the slightly different topology formed by rejoining some of the links (previously cut to get to the reflection covariance) the fringes are allowed to overlap and surround the second periodic orbit formed.  For closed systems without time reversal symmetry we then rejoin more links and create a third (and fourth) periodic orbit.  Also this relaxes a restriction and leads to a modification of the results obtained for the spectral form factor in \cite{Rah}. By including all possibilities we showed that our semiclassical result agrees with the field-theoretical prediction from \cite{Tia}, and hence provides some justification for the phenomenological treatment of effective RMT.

These results are a first step towards the semiclassical calculation of transport and spectral properties including the Ehrenfest time dependence to arbitrary high orders. In the case of transport we found that the second order quantum correction is independent of the Ehrenfest time, whereas the first quantum correction turned out to be proportional to ${\rm e}^{-\tau_E/\tau_D}$ \cite{Ada}: it seems thus not yet possible to identify a general pattern behind the Ehrenfest time dependence of the total contribution to this transport quantity at different orders.  For the related case of the spectral form factor 
the end result seems to possess a simple structure: the contribution from each discrete diagram (i.e.\ those considered without the Ehrenfest time dependence in \cite{Mul1}) to the spectral form factor $K(t)$ is given by a function depending on $t-n\tau_E$ where $n$ is the number of encounter stretches of the underlying diagram. This is the structure predicted by effective RMT and the next step here would be to see if this indeed holds semiclassically to arbitrarily high order.  In principle this would require a general treatment of arbitrarily sized periodic orbit encounters and their fringes and we have the main ingredients \cite{Bro,Bro1} but lack the diagrammatic rules, that are available when the Ehrenfest time vanishes \cite{Heu1,Mul2}, for treating all possible correlations occurring  mainly during fringes.  We wonder if it would be possible to partition the semiclassical diagrams in an efficient way so that any such structure becomes clear, as was the case for the related problem of the correlator of $2n$ scattering matrices \cite{Wal2}.

\section{Acknowledgements}
We thank Cyril Petitjean and Klaus Richter for stimulating discussions and the referees for valuable suggestions.
Financial support by the Deutsche Forschungsgemeinschaft within 
GRK 638 (DW, JK) and by the Alexander von Humboldt Foundation (JK) 
is gratefully acknowledged.

\section{Appendix}

In this appendix we want to calculate the integrals over the stable and unstable coordinates $s$ and $u$ occurring for encounters with periodic orbits for general prefactors $a,b,d$ in front of $t_{{\rm enc},1}+t_{{\rm enc},2}$, $t_s+t_u$ and $\tau_P$, respectively. We therefore consider the expression $J$ defined as
\begin{eqnarray}
\label{ap1}
J&\equiv&\left\langle\int_0^\infty d\tau_p\int_{-c}^c d^2{s}d^2{u} \int_{0}^{\tau_p} dt'\frac{1}{\Omega^2 t_{{\rm enc},1}t_{{\rm enc},2}}\right.\nonumber\\&& \left.\times\exp\left(a\left(t_{{\rm enc},1}+t_{{\rm enc},2}\right)+b\left( t_s+t_u\right)-d\tau_p\right)\right.\nonumber\\&&\left. \times\exp\left(\frac{i}{\hbar}\sum_{i=1}^2{s_i}{u_i} \right)\right\rangle
\end{eqnarray}
with the duration of the fringes before and after the considered orbit approaches the central periodic orbit, $t_s$ and $t_u$, respectively.
The $s,u$- and the $t'$-integrals in $J$ are evaluated in \cite{Bro} yielding
\begin{eqnarray}
\label{ap2}
J&=&\int_0^\infty d\tau_P\frac{a^2}{T_H^2}\tau_P{\rm e}^{-\left(d\tau_P-2a\tau_E\right)}\nonumber\\&&+\int_0^\infty d\tau_P\frac{b^2{\rm e}^{-d\tau_P}}{2T_H^2\left(a-b\right)}\left({\rm e}^{2a\tau_E}-{\rm e}^{2b\tau_E}\right).  
\end{eqnarray}
Performing the $\tau_P$-integral finally yields
\begin{eqnarray}
\label{ap3}
J&=&\frac{a^2}{d^2T_H^2}{\rm e}^{2a\tau_E}+\frac{b^2}{2T_H^2d\left(a-b\right)}\left({\rm e}^{2a\tau_E}-{\rm e}^{2b\tau_E}\right),  
\end{eqnarray}
which is frequently used in the main part of this article.

Furthermore we want to consider a 3-encounter and calculate with arbitrary prefactors $f$ and $g$ in front of $t_{{\rm enc}}$ and $t_s+t_u$ respectively
\begin{eqnarray}
\label{ap4}
K&\equiv&\left\langle\int_{-c}^c d^2{s}d^2{u}\frac{1}{\Omega^2 t_{{\rm enc}}}\exp\left(ft_{{\rm enc}}+g_1t_s+g_2t_u \right)\right.  \nonumber\\  &&\left.\times\exp\left(\frac{i}{\hbar}\sum_{i=1}^2{s_i}{u_i}\right)\right\rangle.
\end{eqnarray}
The first contribution $K_1$ is obtained by setting $g=0$, it yields \cite{Bro1}
\begin{equation}
\label{ap5}
K_1=\frac{f}{T_H^2}{\rm e}^{f\tau_E}.
\end{equation}
The second contribution $K_2$ given by $K-K_1$ is given for $g_1=g_2\equiv g$ by \cite{Bro1}
\begin{equation}
\label{ap6}
K_2=3\frac{g^2}{(2g-f)T_H^2}\left({\rm e}^{2g\tau_E}-{\rm e}^{f\tau_E}\right).
\end{equation}
For $g_1\neq g_2$ we have instead \cite{Bro1}
\begin{equation}
\label{ap7}
K_2=3\frac{g_1g_2}{(g_1+g_2-f)T_H^2}\left({\rm e}^{\left(g_1+g_2\right)\tau_E}-{\rm e}^{f\tau_E}\right).
\end{equation}

\end{document}